\documentclass[]{aa}

\usepackage{graphicx}
\usepackage{natbib}
\usepackage{scalerel}

\usepackage[table]{xcolor}

\usepackage{txfonts}
\usepackage[pdfencoding=auto,psdextra]{hyperref}
\hypersetup{
    colorlinks=true,
    linkcolor=blue,
    filecolor=magenta,      
    urlcolor=blue,
    citecolor=blue
}
\makeatletter
\renewcommand*\aa@pageof{, page \thepage{} of \pageref*{LastPage}}
\makeatother

\usepackage[utf8]{inputenc}

\usepackage[switch, modulo]{lineno}

\begin{document}

\title{Fast emulation of cosmological density fields based on dimensionality reduction and supervised machine-learning}

\author{Miguel Concei\c c\~ao
          \inst{1,2},
          Alberto Krone-Martins
          \inst{3,4}, 
          Antonio da Silva
          \inst{1,2},
          \and
          Ángeles Moliné
          \inst{5}}

\institute{Departamento de F\'{\i}sica, Faculdade de Ci\^encias, Universidade de Lisboa, Campo Grande, 1749-016 Lisboa, Portugal  
              \and
              Instituto de Astrof\'{\i}sica e Ci\^encias do Espac\c co, Faculdade de Ci\^encias, Universidade de Lisboa, Campo Grande, 1749-016 Lisboa, Portugal
              \and
              Donald Bren School of Information and Computer Sciences, University of California, Irvine, 
         Irvine CA 92697, USA 
              \and
  	     CENTRA, Faculdade de Ci\^encias, Universidade de Lisboa, 1749-016 Lisboa, Portugal  
              \and
              Departamento de F\'{\i}sica, ETSI Navales, Universidad Polit\'ecnica de Madrid, Avda. de la Memoria, 4, 28040 Madrid, Spain
                 }

%
%
\abstract{
N-body simulations are the most powerful method to study the non-linear evolution of large-scale structure. However, they require large amounts of computational resources, making unfeasible their direct adoption in scenarios that require broad explorations of parameter spaces. In this work we show that it is possible to perform fast dark matter density field emulations with competitive accuracy using simple machine-learning approaches. We build an emulator based on dimensionality reduction and machine learning regression combining simple Principal Component Analysis and supervised learning methods. For the estimations with a single free parameter, we train on the dark matter density parameter, $\Omega_m$, while for emulations with two free parameters, we train on a range of $\Omega_m$ and redshift. The method first adopts a projection of a grid of simulations on a given basis; then, a machine learning regression is trained on this projected grid. Finally, new density cubes for different cosmological parameters can be estimated without relying directly on new N-body simulations by predicting and de-projecting the basis coefficients. We show that the proposed emulator can generate density cubes at non-linear cosmological scales with density distributions within a few percent compared to the corresponding N-body simulations. The method enables gains of three orders of magnitude in CPU run times compared to performing a full N-body simulation while reproducing the power spectrum and bispectrum within $\sim 1\%$ and $\sim 3\%$, respectively, for the single free parameter emulation and $\sim 5\%$ and $\sim 15\%$ for two free parameters. This can significantly accelerate the generation of density cubes for a wide variety of cosmological models, opening the doors to previously unfeasible applications, as parameter and model inferences at full survey scales as the ESA/NASA Euclid mission.
    }
%
%
\keywords{Cosmology: large-scale structure of Universe; Methods: numerical; N-body simulations; Machine Learning}
%
%
\titlerunning{Fast emulation of cosmological density fields}
\authorrunning{Concei\c c\~ao, M., et al.}
\maketitle
%
%
%
%

\section{Introduction}
Numerical N-body simulations are arguably the most powerful method to describe the formation and evolution of cosmological structure, especially at late times when the growth of perturbations, driven by gravitational collapse, becomes highly non-linear. They provide high-accuracy predictions on how the dark matter (DM) density and velocity fields evolve, allowing the identification of model signatures and observables that can be compared against observations. 
This process often requires the generation of a very large number of computationally intensive simulation runs, posing a prohibitive bottleneck on the number of simulated models that can be explored in the preparation and scientific exploitation of modern cosmological surveys.
This penalizes model inference studies, specially using Bayesian methods considering the shear data volume of modern sky-surveys such as ESA/Gaia \citep{Gaia2016, 2022arXiv220800211G}, ESA/Euclid \citep{EuclidM, ECSurveyI2022} and time-resolved surveys as Zwicky Transient Facility \citep{ZTF} and LSST/Rubin \citep{VeraRub}.

One typical way to address these types of challenges is through the adoption of simulation emulators. These emulators try to reproduce the output of a simulator by the adoption of multiple regression methodologies \citep[e.g.,][]{OhaganKingman}. Gaussian Processes, for instance, have been used in such scenarios \citep[e.g.,][]{Currin88, Currin91}. More recently, the attention has been shifted to the Machine Learning (ML) approach, with Deep Models populating most of the density field emulation landscape.

Emulation of dark matter density fields with deep learning techniques such as {\it Generative Adversarial Networks} (GANs) or {\it Variational Auto-Encoders} (VAEs) methods have been proposed, both in 2D \citep[e.g.,][]{DML1, DML3} and 3D \citep[e.g.,][]{DML2} and showed that it is possible to use Deep Learning to emulate DM density fields using the Lagrangian approach \citep[e.g.,][]{DML4, giusarma2019, DML6, ramanah2020, 2022arXiv220604594J}, instead of the less general Eulerian approach. 
The emerging picture is that Deep learning models tend to efficiently emulate DM fields, but generally require large training sets. Their accuracy starts to degrade below scales $k\simeq 1 h\, {\rm Mpc}^{-1}$, i.e. mildly inside the strong non-linear regime of structure formation.
Other works \citep[e.g.,][]{fpcaP,upscP},  focused on the emulation  and resolution enhancement of N-body simulations using these methods with minimal computational resources.

Although the Deep Learning approach enables a decrease in time and computational resources needed to obtain an output similar to that of an N-body simulation, it still remains a reasonably complex and hardly interpretable approach. Of course, since the N-body density fields usually consist in extremely large arrays, containing millions of matter density values with a fair degree of non-linear behaviour, it is compelling to think that only deep models should be able to handle such data. However, considering that we are dealing with non-random data with an high degree of symmetries brought by the cosmological principle, it should be possible to take advantage of the redundancy in this data, compressing it in a more compact basis. Using this reduced data representation, it becomes more feasible to apply simpler machine learning models, decreasing further the amount of computational resources and time needed to obtain the outputs, and hopefully increasing the interpretability of the models. 

In this work, we propose a methodology that aims to achieve such goals. Our methodology first adopts a simple data projection method, Principal Component Analysis (PCA) \citep{Pearson, Hotelling, Jollife}, to project the N-body training data into a smaller basis of coefficients called {\it Principal Components} (PCs), and then uses the transformed representation as an input to a set of simple and interpretable supervised learning methods: Random Forest \citep[RF;][]{RF}, Extremely Randomized Trees \citep[ERT;][]{ERT}, shallow Neural Networks \citep[NN;][]{ANNs} and Support Vector Machines \citep[SVM;][]{SVM}. 

We test our methodology in two different scenarios. In the first scenario, we perform the emulations using the dark matter density ($\Omega_m$) as a single free-parameter. In the second scenario, we slightly increase our dataset and include the redshift, $z$, as an additional free-parameter, enabling us to test the generalization capacity of our models, under higher dimensional contexts. To gauge the performance of the trained models and of our proposed methodology in general, we compare the Power Spectrum and Bispectrum of our emulations with that of an N-body simulation generated for the same cosmology and initial conditions. In this way, we ensure that our emulated density fields have their statistical properties as close to those of the true simulation as possible, at large and small scales.

In this paper we present our proposed methodology and the adopted comparison metrics in Sec. \ref{sec:methodology}. Afterward, we show the results of its application to emulate N-body simulation density fields in Sec. \ref{sec:results}. Then, we present a small discussion, with the computational efficiency of our method addressed in Sec. \ref{sec:disc}. Finally we present the conclusions in Sec. \ref{sec:conc}.

\section{Methodology}
\label{sec:methodology}
A consequence of the highly redundant amount of information present within the large scale structure of the Universe, is that this information can be highly compressible. Accordingly, the density field emulator design described in this section is based first on the application of a dimensionality reduction method and then on a supervised machine-learning method. These methods requires the construction of a training (and test) set of N-body simulations and evaluation metrics that we will also describe in this section.

\subsection{The density field emulator}
\label{sect:estimator}

\subsubsection{PCA projection}
We start from a library of $k$ density fields, each one of them created from an independent N-body simulation with parameters $\mathbf{D}_k$. Each cube of size $n^3$ of this library is rearranged into a row of a matrix $\mathbf{M}_{k\times n^3}$, with $k$ rows by $n^3$ columns. The first step in our methodology is to solve a data compression problem: to find a sparser representation $\mathbf{S}_{k\times l}$ of the matrix $\mathbf{M}$, with $l \lll n$ and $l \le k$. We apply PCA, which is a very simple and well-established methodology adopted in multiple contexts in Astronomy \citep[e.g.][]{2009MNRAS.395...64S, 2011A&A...527A..49I,2014A&A...561A..57K,ECEmulator2IX2021,fpcaP,upscP} to perform this compression. We note, though, that although we adopt PCA here for simplicity, any other invertible dimensionality reduction and/or compression methods could be used. To perform the PCA we adopt a Singular Value Decomposition. It decomposes the $\mathbf{M}$ matrix in three matrices $\mathbf{M=UZV^{T}}$ where $\mathbf{U}$ is an unitary $k\times k$ matrix,
$\mathbf{Z}$ is a diagonal matrix with the singular values and $\mathbf{V}$ is a $n^{3} \times k$ matrix containing the eigenvectors of the covariance matrix in each column -- the so called the principal directions. The principal componenents (e.g $\mathbf{MV}$, or equivalently $\mathbf{UZ}$) are obtained by projecting the data in the principal directions. The crucial step to perform the dimensionality reduction is to select the first $l$ columns of $\mathbf{U}$ and the left-right $l\times l$ block of $\mathbf{Z}$ their product gives us our $\mathbf{S}_{k\times l}$ matrix with the first $l$ principal components, capturing most of the data variance. The reconstruction or decompression can be performed with the first $l$ PC's by  multiplying $\mathbf{S}$ by the transpose of the matrix $\mathbf{V}$, $\mathbf{V^{T}}$.

\subsubsection{Supervised learning regression}
The second step of our methodology is to solve a regression problem, i.e., to find a map $f := \mathbf{D} \to \mathbf{S}$. And so, given any set $i$ of universe parameters $\mathbf{D}_i$, it is possible to estimate the compressed field $\mathbf{S}_{i}$, that after decompression generates an estimation of the density field $\mathbf{M}_{i}$ as if we would have executed a full N-body simulation with the parameters $\mathbf{D}_i$. 
To construct this map, we studied four supervised machine learning algorithms in this work: Extremely Randomized Trees, Random Forests, Neural Networks and Support Vector Machines. Our motivation to use multiple methods is to compare the emulation yields obtained with them in this first exploratory study, since they could, in principle, present better results in different subdomains of the parameter space as well as in different scales of the density field. That could justify, in a future study, the construction of more precise ensembles \citep[e.g.][]{10.1145/2379776.2379786}.

The adoption of supervised learning methods in astronomy has a long history \citep[e.g.][]{2022arXiv221201493D}. Neural Networks \citep{ANNs} and Support Vector Machines \citep{SVM} are classical methods that have been successfully used in multiple contexts. For instance, before the recent boom in citations to computer science works from astronomical papers \citep{2022RNAAS...6..113V}, mostly due to the spread of deep learning approaches, Neural Networks have been used for almost four decades in astronomy already. Their earlier adoption were in optimization \citep[][]{1986ApJ...310..473J}, in heliophysics \citep[][]{1990EOSTr..71..677K}, adaptive optics \citep[][]{1990Natur.348..221A, 1991Natur.351..300S}, star/galaxy separation \cite[][]{1992AJ....103..318O} and galaxy classification \citep[][]{1992MNRAS.259P...8S}. Support Vector Machines on the other hand, started to be used almost exclusively in extragalactic contexts in the 2000s \citep[e.g.][]{2002SPIE.4847..371Z, 2005PASP..117...79W,2007A&A...470..761T, 2008A&A...478..971H, 2008AIPC.1082..151K,2008MNRAS.391.1838B}, but they have also found adoption in classification of transients \citep[][]{2008AIPC.1082..287M}, in stellar classification \citep[e.g.][]{2010A&A...522A..88S}, interstellar medium structure \citep[][]{2011ApJ...741...14B}, heliophysics \citep[e.g.][]{2012ApJ...746...12A}, among others.

Random Forests \citep[][]{Ho:1995:RDF:844379.844681,RF} started to be used in astronomy in the early 2000s. As SVMs, they were first used in an extragalactic context, to search for quasars \citep{2003sca..book..243B} and supernovae \citep{2007ApJ...665.1246B}, to identify high-energy sources \citep{2008AIPC.1082..307S} and to perform photometric redshift estimation \citep{2010ApJ...712..511C}, but they later found other applications such as variable star classification \citep{2011ApJ...733...10R, 2011MNRAS.414.2602D}, feature selection for stellar membership analysis \citep{2014A&A...563A..45S}, anomaly detection \citep{2014ApJ...793...23N} and young stellar object classification including missing data \citep{2017A&A...597A..90D}. Finally, the last method adopted here, Extremely Randomized Trees \citep{ERT}, is a supervised learning method that is a variant of the more commonly used Random Forests. Although not widely adopted in astronomy as the other methods, in the past decade ERTs have been successfully used in the context of stellar astrophysical parameter determinations \citep{2013A&A...559A..74B, 2018MNRAS.473.1785D}, quasar selection \citep{2014MNRAS.439..703G}, prediction of galaxy properties from dark matter halo simulations \citep{2016MNRAS.455..642K} and to discover new gravitationally lensed quasars \citep{2018A&A...616L..11K, 2019A&A...622A.165D}. As RFs, ERTs are based on ensembles of decision trees, but they split trees differently: while RFs split trees deterministically, ERTs splits are chosen entirely at random \citep{ERT}.

\subsubsection{Algorithm Optimization}
\label{sec:optimization}

As with most machine learning methods, the aforementioned supervised learning algorithms depend on hyper-parameters that can be optimized. 
For each method, balancing hyper-parameters can be important to improve the speed and performance of the method. 
To optimize them we explore each method's hyper-parameter space with different optimization schemes. We chose to experiment with the standard Grid-Search approach \citep[e.g.][]{726791,Bergstra:2012:RSH:2503308.2188395}, Bootstrap Aggregating \citep[e.g.][]{bootstrap} and K-fold Cross-Validation \citep[K-CV; e.g.][]{hastie2009elements}, optimization methods. 
Each optimization method provides a different optimal model. To discriminate between them we compare the best emulation from each method with the corresponding, \textit{ground truth}, test simulation using the cosmological power spectrum, see Sec.~\ref{sec:emul_stats}. 
For this, we define the power spectrum distance ratio metric as:
\begin{equation}
    D=\sum_{i=1}^{n} \Big |1-\frac{P^i_{est}}{P^i_{sim}}\Big |
\end{equation}
where $n$ is the power spectrum vector size, $P^i_{est}$ and $P^i_{sim}$ are the estimated and simulated power spectrum of the ith wave-mode, $k_i$, respectively.
Consequently, we take the best performing optimization method, as the one providing the smallest $D$. 

The application of the these methods to the different supervised learning variants of our emulator involves optimizations over different sets of hyper-parameters. Our choice of hyper-parameters and methodology, in each case, was as follows.

For Random Forest, the hyper-parameter optimization was on the number of trees/estimators in the ensemble ({\tt ntree}), the minimum number of observations allowed in the terminal leaves of the trees ({\tt nodesize}) and the number of independent variables randomly sampled as candidates for each split ({\tt mtry}). Therefore, the optimization of the latter parameter, is only possible for regression problems with more that one free parameter (i.e., in the case of emulations in $\Omega_m$ and $z$).  
To perform the search over this hyper-parameter space, we first did a simple grid-search over the number of trees for the values ntree=(500, 1000, 2000), fixed the best performing value and proceeded to optimize the remaining hyper-parameters, experimenting with both Bootstrap and K-fold Cross Validation. This procedure was implemented using the {\tt e1071} package \citep{e1071}.

For the Extremely Randomized Trees, our optimization approach is similar to the RF case, except that instead of optimizing {\tt nodesize} we optimized {\tt numRandomCuts}, which as the name implies, is defined as the number of random splits considered at each partitioning node of the tree. Then the optimization proceeded in similar way to that of RF. First, by fixing {\tt ntree} with a Grid-Search over the same possibilities, followed by a search over {\tt mtry} and {\tt numRandomCuts} and experimenting with the same optimization schemes. This procedure was implemented using the {\tt caret} package \citep{caret}.   

Regarding the Neural Networks we optimized the parameter {\tt size}, corresponding to the number of nodes in the single hidden layer and the parameter {\tt decay}, corresponding to a regularization constant in the weight decay formula, which constrains the number of free-parameters (weights) in the neural network model.

Finally, for the Support Vector Machines we first did a Grid-Search over the different kernel possibilities, with the remaining hyper-parameters set to their default values. After finding the optimal kernel we proceeded to optimize the hyper-parameters {\tt cost} and {\tt gamma}.

\subsection{Training and Test Data set}

One advantage of our emulation method is that it is a simple and generalizable approach to the emulation of N-body density fields. To demonstrate the flexibility of the method we decided to make all computations on a simple computer system (a quad-core Intel core i5 with 16Gb of RAM and no discrete GPUs), running Linux. 
We therefore planned our set of N-body simulation runs, 
that we will use for the training of our emulator, with size and resolution specifications capable of resolving a range of non-linear evolving cosmological scales in the chosen platform. 

Our simulations were generated for a family of cosmological models based on the flat, Cold Dark Matter, $\Lambda${\it-CDM} cosmology with Hubble parameter $h=0.7$ and dark matter density parameter $\Omega_{dm}=\Omega_m=1-\Omega_{\Lambda}$, where $\Omega_m$ is the total matter parameter and $\Omega_{\Lambda}$ is the energy density parameter associated with the cosmological constant. In this way the energy density of radiation is neglected and baryons are treated in the simulations as collisionless dark matter. 
The initial conditions (IC) matter power spectrum was modeled with the {\it CDM} BBKS transfer function with a shape parameter given by $\Gamma =\Omega_{m}\, h$ \citep{BBKS_1986, Sugiyama_1995} and a normalization controlled by the $\sigma_8$ parameter, which is the square root of the variance of the smoothed over-density field on scales of $8\,h^{-1}\textrm{Mpc}$ at present. The primordial power spectrum index of scalar perturbations was set to $n_s=1$.
The simulation runs were generated with the {\tt Hydra} public code \citep{Hydra}, that implements an adaptive mesh refinement AP$^3$M method to compute gravitational forces and evolves a set of DM mass particles in serial mode. 
We used the {\tt cosmic} initial conditions generator provided in the Hydra code package to generate fields for $160^3$ particles of dark matter in a cubic volume with comoving length $L = 100\,h^{-1}$ Mpc on the side. 
In physical coordinates, the simulation's gravitational softening was held fixed to 
$25h^{-1}\,$kpc up to $z = 1$, and scaled as 
$50(1+z)^{-1}h^{-1}\,$kpc 
above this redshift. 
With these choice of parameters and force resolution, our simulations could in principle resolve scales, in high density regions, within a minimum and maximum wave-mode of about $k_{min}\simeq 0.06 \, h\,{\rm Mpc}^{-1}$ and $k_{max}\simeq 30\,h\,{\rm Mpc}^{-1}$, respectively. 
All our simulation runs share the same initial conditions seed and start from an initial snapshot realization at z = 49 that is evolved until z = 0. 

Our template of simulation runs consists of 23 simulations with $\Omega_m$ taking values in the interval $\Omega_m \in [0.05,0.6]$ with a regular separation equal to $\Delta \Omega_m = 0.025$. All other parameters were held fixed with the exception of the primordial shape parameter that varied according to $\Gamma=\Omega_m h$. For each of these runs we store 10 snapshots (cubic volumes) containig particle positions and velocities at each redshift. 
In addition to these data we run a simulation with $\Omega_m=0.31$ and the same parameter setup, to be used use as a {\it ground truth} run that is not part of our training set.
Before using all these runs with our emulation pipeline we first 
need to sample the dark matter densities on a regular cubic grid for all snapshots. In fact, {\tt Hydra} is a particle based code that follows volume elements (the particles) using a {\it lagrangean} description, so it does not provide densities directly. To convert these into a {\it eulerian} grid of densities, we adopted a mass-assignment scheme that treats dark matter as SPH \citep{lucy1977,monagham1992} particles. 
For this, we use the {\tt darkdens} code in {\tt Hydra}, to compute smoothing lengths, and a fast SPH grid method based in the mapping algorithm in \citep{dasilva2001, elsa2012}.
For the purpose of this work, we adopted a regular grid with $128^3$ voxels inside the simulation volume, which is grid choice that limits the resolution in our emulations to a maximum wave-mode of about $k_{max}\simeq 3\,h\,{\rm Mpc}^{-1}$. Finally, our simulation's training set is therefore a template of simulations cubes, all with $128^3$ voxels, spanning in a range of 23 values of $\Omega_m$ and 10 values of $z$ (which amounts to total of 230 simulation cubes). In practice we will work with different subsets of these data for the cases of emulations with one or two-free parameters, as described below.

\subsection{The Emulation Pipeline}
\label{sec_pipeline}

The first step in our emulation pipeline is to project our training dataset with PCA. We concatenate simulation arrays as vectors containing the base 10 logarithm of the densities cubes. The vectors are then put in a matrix form before applying the PCA method with the data re-scaled and centered. 

For emulations with a single free parameter (i.e. with $\Omega_m$), this procedure leads to 23 PCs for each simulation. Once the training data is projected in its principle component representation
we use the PCs to build and optimize the supervised learning algorithms. 
Since the density fields are represented by PCs, and we aim to estimate the field itself, the training is performed using the PCs as dependent variables and $\Omega_m$’s as independent variables. To achieve this, we create a 23 iteration loop where in each iteration our algorithms build a regression model accounting for that specific PC and the same constant set of $\Omega_{m}$ values.
For the case of the neural networks, the data is 
re-scaled to the [0,1] range, to reduce the chances of getting stuck in local minima.
After having the new PCs, the final estimated cube is obtained by de-projecting them on to the original representation using the inverse de-projection operations:
\begin{equation}\label{eq:deproj}
{\bf M_{orig}}=10^{{\bf S}{\bf L^T}{\bf s^T}+{\bf c^T}}
\end{equation}
where ${\mathrm{\bf S}}$, ${\mathrm{\bf L}}$, ${\mathrm{\bf s}}$ and ${\mathrm{\bf c}}$, are the scores, loadings, scaling and centering matrices and vectors, respectively.

For emulations with two parameters, the main difference with the one-free parameter case is the size of training data set. To incorporate the training in redshift for the chosen computer platform we decided to use only four redshifts, z = [0, 0.25, 0.75, 1], from our initial dataset. 
For each of these redshifts we keep the same 23 $\Omega_{m}$ density cubes we have in the case of regressions with a single parameter, $\Omega_m$. Our training data set therefore has $23\times 4 = 92$ simulation cubes, and we kept the cube with $\Omega_{m}$ = 0.31 and $z = 0.5$ to compare emulation results against the corresponding simulation ground truth.
The whole pipeline takes significantly longer times to run 
in this case. This is because our data set is significantly larger as it involves working with a basis 92 PCs, instead of only 23 for the single free parameter emulation.  
Our emulation pipeline was fully written in the {\tt R} programming language \citep{Rman}. This includes the module of PCA decomposition, implementation of the different supervised regression algorithms, optimization with the Grid-Search, Bootstrap Aggregating and K-fold Cross-Validation methods, and finally density cube emulations using up to two-free parameters.

\subsection{Evaluation statistics }
\label{sec:emul_stats}
To assess the performance of our emulation method we compare, in Sec.~\ref{sec:results}, the power spectrum and bispectrum of our emulated volumes with those of the ground truth simulations. The power spectrum statistic is also used as part of our optimization of the method, as described in Sec.~\ref{sec:optimization}.

The power spectrum is the Fourier Transform of the 2-point correlation function:
\begin{equation}
    \xi(r,z)=\frac{V}{(2\pi)^3}\int \langle |\delta_k(\vec{k}, z)|^2\rangle e^{-i\vec{k}\cdot \vec{r}}d^3k
\end{equation}
where $V$ is the box volume and 
\begin{equation}
    P(k,z)=\langle |\delta_k(\vec{k}, z)|^2\rangle = \frac{1}{V}\int \xi(r, z)e^{i\vec{k}\cdot\vec{r}}d^3r
\end{equation}
is the power spectrum. For a Gaussian distribution of densities, $P(k,z)$ provides a complete statistical description of the density field. However, the density field in N-body simulations in not Gaussian due to the non-linear growth of cosmic structure. We compute the non-linear power spectrum of the emulations and ground truth simulations with the public version of the {\tt powmes} code \citep{powmes}.

To measure departures from non-gaussianity, we also compute the Bispectrum of the density cubes. The Bispectrum is associated with the Fourier Transform of the 3-point correlation function and, as an higher order correlation function, it is sensitive to the non-linear growth of cosmic structure. It is usually defined as {\it B}, see e.g. \cite{Sef2005}:
\begin{equation}
 \langle\delta_{\vec{k}_1}\delta_{\vec{k}_3}\delta_{\vec{k}_3}\rangle = \delta_D(\vec{k}_1+\vec{k}_2+\vec{k}_3)B(k_1,k_2,k_3)
\end{equation}
where $\delta_D$, the Kronecker delta function, imposes a triangular relation between wave-mode vectors, $\vec{k}_1+\vec{k}_2=-\vec{k}_3$. Here we adopt a configuration of wave-modes as in \citep{ramanah2020, giusarma2019} where $\vec{k}_1$ and $\vec{k}_2$ have fixed magnitudes, $k_1=|\vec{k}_1|=0.15h\,{\rm Mpc}^{-1}$ and $k_2=|\vec{k}_2|=0.25h\,{\rm Mpc}^{-1}$, and $k_3=|\vec{k}_3|$ varies with the angle, $\theta$, between $\vec{k}_1$ and $\vec{k}_2$. This choice of wave-mode amplitudes are well inside the wave-mode window of our simulations, and they typically correspond to the smallest scales probed by galaxy clustering power spectrum determinations \citep{giusarma2019}. In this work we use the {\it Pylians3} public software\footnote{https://github.com/franciscovillaescusa/Pylians3} to compute the Bispectra presented in Sec.~\ref{sec:results}.  

\section{Results}
\label{sec:results}
In this Section we present our main results. We show how each of our estimated density fields compare against the simulated density field using their power spectra and bispectra as comparison metrics. For the single free-parameter case, we present the results concerning the implementation of our pipeline at four different redshifts.

\subsection{The Single Free-Parameter Case}
\begin{figure}
	\centering
	\includegraphics[width=\columnwidth]{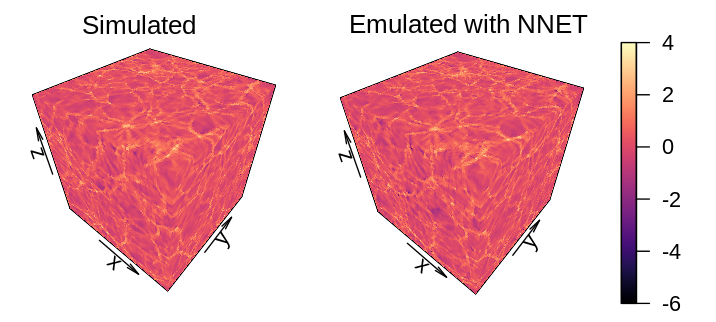}
	\caption[Visual Representations] {Visual Representation of the Ground Truth Simulation (left) against our Neural Network Emulator (right) at redshift $z=0$.}
	\label{fig:predVSsim}
	\centering
\end{figure}

For emulations with a single free-parameter, we first show a visual representation of one of our emulations (the best performing one) with the actual Ground Truth simulation in Fig.~\ref{fig:predVSsim}, where it is possible to notice a remarkable similarity between both representations.

\begin{figure}
	\centering
	\includegraphics[width=\columnwidth]{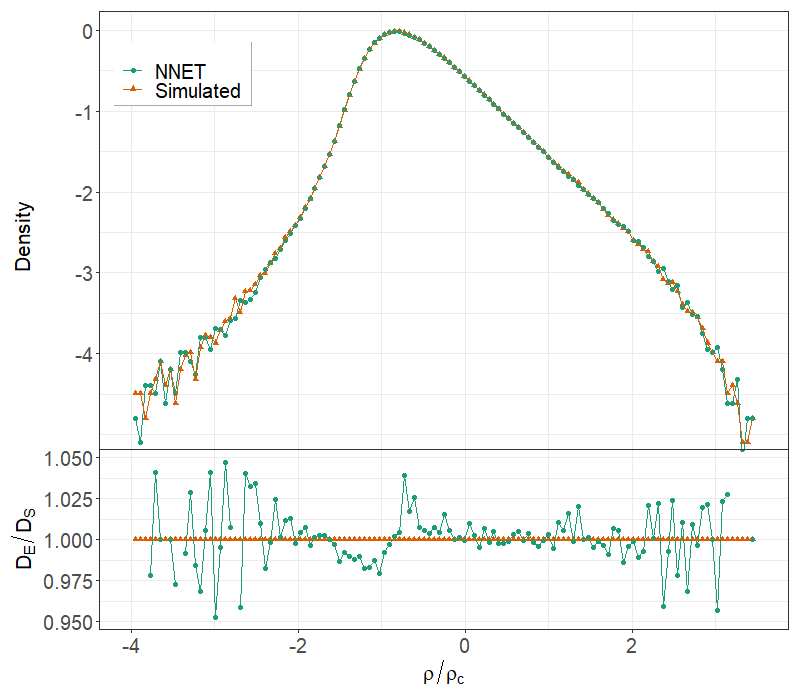}
	\caption[Density distributions] {Top panel: Distribution of the base-10 logarithm of voxel over-densities in the ground truth simulation and emulation volume represented in Fig.~\ref{fig:predVSsim}. Bottom panel: ratio between the distribution functions displayed in the top panel. The ratio $\Delta=\rho/<\rho>$ is the over-density contrast.}
	\label{fig:density_hist}
	\centering
\end{figure}

Moving now towards a more quantitative description, we show in Fig.~\ref{fig:density_hist} the distribution of the base-10 logarithm of the voxel densities in the emulated and ground truth volumes represented in Fig.~\ref{fig:predVSsim}.

The densities are in units of the mean background density, i.e. are voxel over-densities, $\Delta=\rho/<\rho>$. The distributions (in the top panel) are clearly non-Gaussian and their ratios (displayed in the bottom panel) indicate that the emulator performs best in the range of densities that are more frequent in the simulated volume. This result confirms expectations because the tails of the distributions are rare over/under-densities (in high density strutures or deep voids) that are also under-represented in the training dataset. Nevertheless the agreement between the distributions is remarkable, especially if we take into account that this emulation was done with a training data set with only 23 density cubes and a Neural Network geometry with a single hidden layer. Differences are typically bellow 2.5\% in the range $0.01 <\Delta<100$. Outside this range, where densities correspond to voxels inside collapsed halos and deep voids, the agreement between distributions is still better than 5\%.

\begin{figure*}
	\centering
	\includegraphics[width=\textwidth]{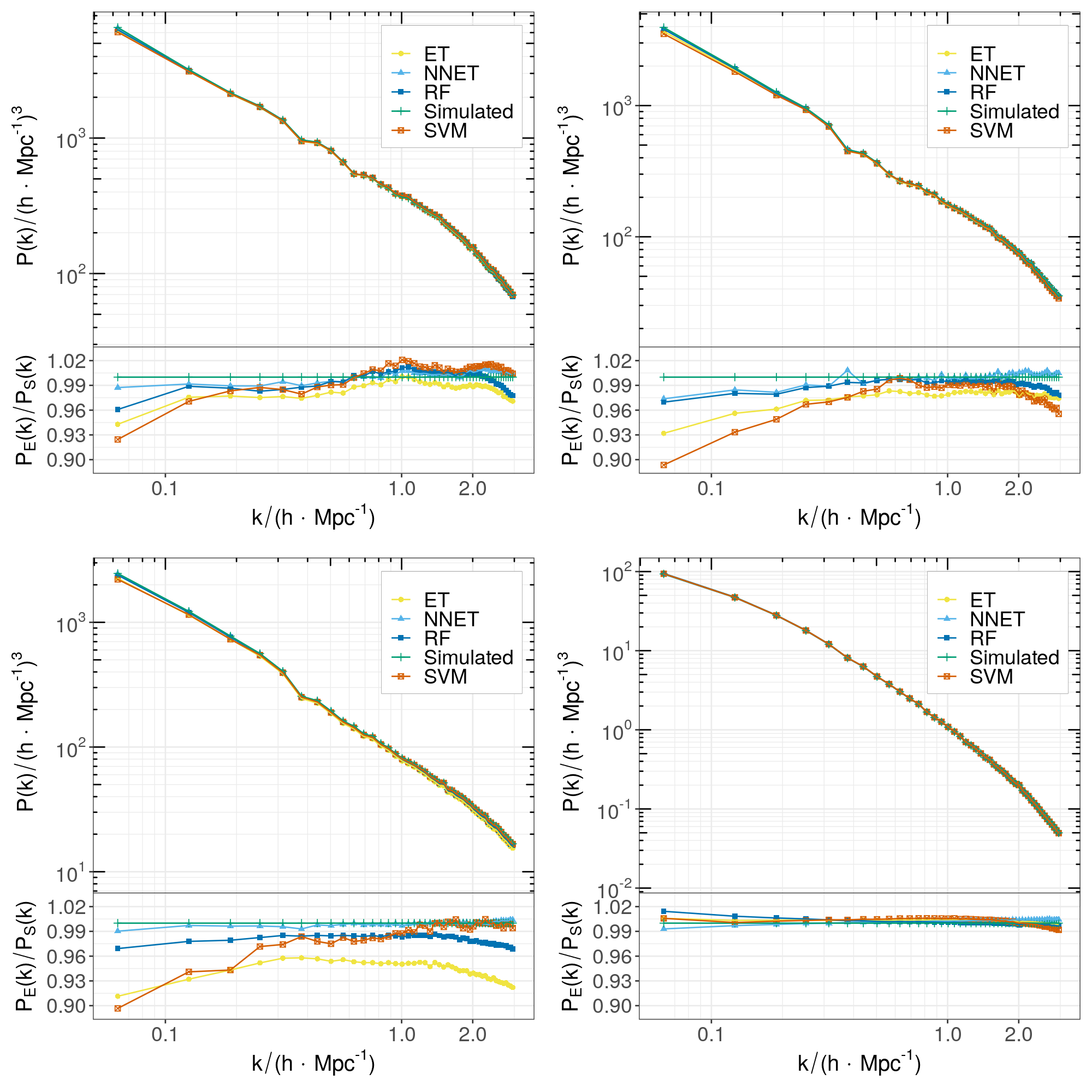}
	\caption[Power Spectra of the Estimated Density Fields (1 free-parameter case)]{Power Spectrum emulations at redshifts $z=0$ (top left), $z=0.5$ (top right), $z=1$ (bottom left) and $z=10$ (bottom right) for model regressions with a single parameter, $\Omega_m=0.3$. The top panel of each plot shows the power spectra obtained with our different emulation variants, Extremely Random Trees (ET), Neural Networks (NNET), Random Forest (RF), and Support Vector Machines (SVM). The bottom panel in each plot shows the ratio of the different power to the power of the true N-body simulations' power spectrum ($P_E(k)=$ Estimated;$P_S(k)=$ Simulated).  
    }
	\label{fig:PowerAll1p}
	\centering
\end{figure*}

Fig.~\ref{fig:PowerAll1p} shows the power spectrum from emulations with one free-parameter, $\Omega_m=0.3$, at fixed redshifts, $z=0$ (top left), $z=0.5$ (top right), $z=1$ (bottom left) and $z=10$ (bottom right). In each plot, the top panel shows the power spectra of the ground truth simulation and ouremulator implementations, featuring Extremely Random Trees (ET), Neural Networks (NNET), Random Forest (RF), and Support Vector Machines (SVM) as supervised learning algorithms. The bottom panel in each plot shows the ratio of the power spectrum of different emulator methods to the the ground truth power spectrum.

Overall, the Neural Networks is our best performing algorithm, achieving an accuracy above 99\% at all redshifts and for most part of the $k$ domain. Random Forest manages to show similar accuracy at all redshifts except at $z=1$ where performance decreases slightly, but still achieves ratios above the 96\% level at all $k$.
The other algorithms have somewhat worse performances, particularly at $z=0.5$ and $z=1$.
At $z=10$ all algorithms show a strong agreement with the power spectrum of the ground truth simulation. This agrees with expectations because most of the resolved scales in our simulations are still evolving linearly at that epoch, and therefore strong, under-represented, over/under-densities do not have enough time to form by that redshift.
In the light of this argument one could expect larger power spectrum deviations as redshift decreases, because progressively larger non-linearities devolop at low $z$. Our results seem to contradict this trend at redshift $z=0.5$ and bellow, because the different power spectrum ratios in Fig~\ref{fig:PowerAll1p}, generally, tend to approach 1 as redshift decreases. But this apparent contradiction can still be understood as a consequence of the level of representation (amount) of non-linearities in the training set, if we take into consideration the dynamics of the $\Lambda$CDM model at $z<1$. Indeed, below this redshift, the cosmological constant reverts the Hubble expansion rate from a decelerated phase, due to matter domination, into an accelerated expansion. As larger over and under-densities develop inside our probed volume, their number is relatively better represented in the training set below this redshift than at higher $z$ (where structure grows in a decelerating matter-dominated universe).  

Another interesting feature of our results is the decrease of power spectrum ratios at small $k$, i.e. at large scales. This trend is common to all algorithms, and again can be interpreted as a lack these independent modes scales in the training data set.

\begin{figure*}
	\centering
	\includegraphics[width=\textwidth]{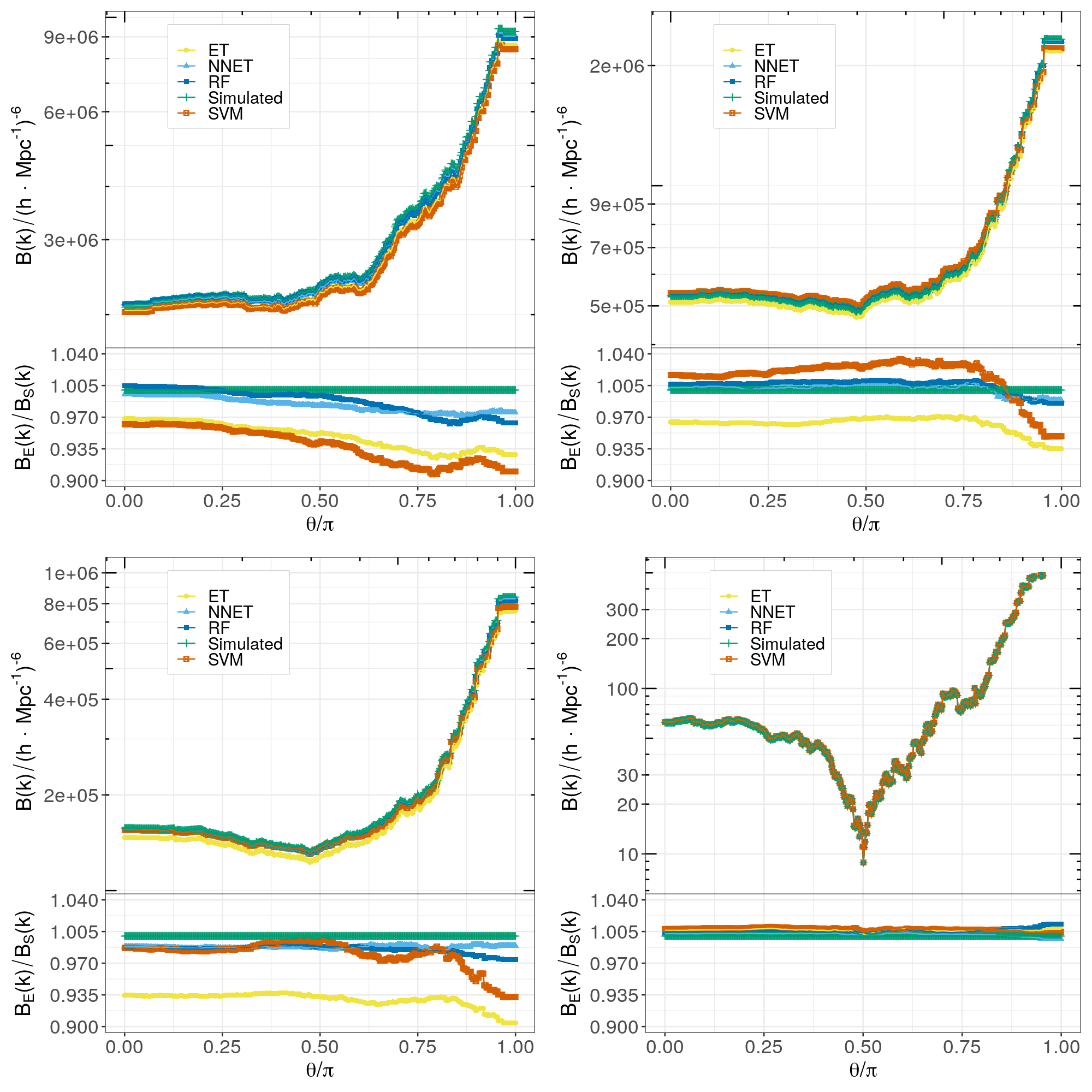}
	\caption[Bispectra of the Estimated Density Fields (1D case)]{ Bipectrum emulations at redshifts $z=0$ (top left), $z=0.5$ (top right), $z=1$ (bottom left) and $z=10$ (bottom right) for model regressions with a single parameter, $\Omega_m=0.31$. The top panel of each plot shows Bispectra for our different emulation variants, Extremely Random Trees (ET), Neural Networks (NNET), Random Forest (RF), and Support Vector Machines (SVM). The bottom panel in each plot shows the Bispectra ratios from Emulations, $B_E(\theta)$, and the $B_S(\theta)$ the ground truth simulation). }
	\label{fig:Bispecallz1p}
	\centering
\end{figure*}

Figure \ref{fig:Bispecallz1p} presents results from our Bispectrum analysis.
As before, the figure shows bispectra and bispetra ratios at redshifts, $z=0$ (top left), $z=0.5$ (top right), $z=1$ (bottom left) and $z=10$ (bottom right), for the $\Omega_m=0.3$ cosmology.
The relative performance of the algorithms is similar to the power spectrum results.
The NNET tends to be the best performing 
algorithm, followed closely by the Random Forest implementation, whereas the other algorithms show lower performances.
Our results also show that at the extremes of the $\theta$ domain the different algorithms also tend to under-perform, especially at large $\theta$. 
The level of disagreement between emulated and ground truth bispectra is also larger than in the case of the power spectrum.
The larger deviations from the ground truth are a consequence of the larger uncertainties, generally, associated with the measurement of tree-point, and other higher-order, correlation functions from a finite volume. 
On the other hand, the larger deviations at large $\theta$ are linked to
the choice of our bispectrum configuration, where $k_1=0.15h\,{\rm Mpc}^{-1}$ and $k_2=0.25h\,{\rm Mpc}^{-1}$ were set to scales that are closer to the lower end of $k$ domain in our simulations and therefore correspond to larger scales that are less represented in our training set. 

\subsection{The Two Free-Parameters Case}
In an attempt to expand the cosmological parameter space of our emulation method and test it under higher dimensional settings, we
included the redshift as an additional free-parameter in our emulation pipeline. 
Here we present results for emulations at $\Omega_m=0.31$ and $z=0.5$

\begin{figure}
	\centering
	\includegraphics[width=\columnwidth]{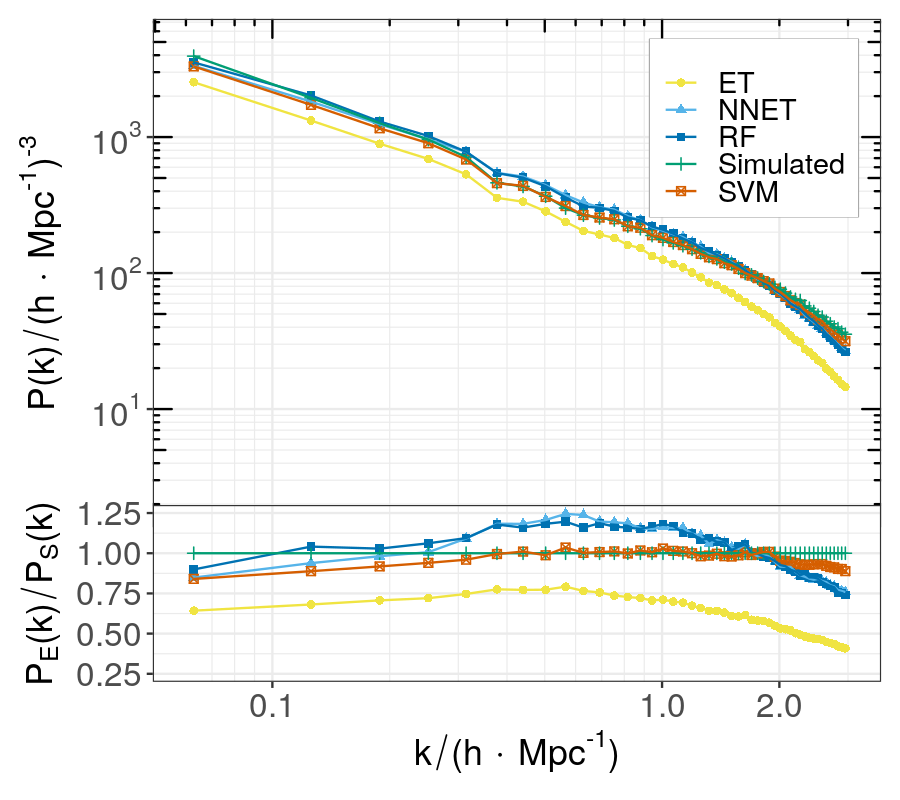}
	\caption[Power Spectra of the Estimated Density Fields (2D case)]{The Power Spectra of the estimated Density Field using the four algorithms on the two free-parameter ($\Omega_m$,z) at $\Omega_m=0.31$ and $z=0.5$.
 Dataset. {\bf Top Panel:} Power Spectra Curves. {\bf Bottom Panel}: Power Spectra Ratio ($P_E(k)=$ Estimated;$P_S(k)=$ Simulated).}
	\label{fig:PowerAll2p}
	\centering
\end{figure}

Fig.~\ref{fig:PowerAll2p} shows the power spectrum (top panel) and power spectrum ratios (bottom panel) of our emulator variants versus the ground truth simulations.  

As expected, there is a general decrease of performance of the algorithms due to the increased dimensionality of the problem.
NNET is no longer the best performing algorithm, with errors reaching $25\%$ at around $k=0.6$ $\textrm{h} \textrm{Mpc}^{-1}$. Both RF and NNET algorithms over-estimate the power spectrum in the range around $0.1 h \,\textrm{Mpc}^{-1}\leq k \leq 2h\, \textrm{Mpc}^{-1}$, while SVM and ET consistently underestimate it.

For our emulation setup with 2 free-parameters, the SVM is by far the best performing algorithm. It achieves an accuracy of more than $95\%$ for $0.3 h \,\textrm{Mpc}^{-1}\leq k \leq 2h\, \textrm{Mpc}^{-1}$, decreasing to about $90\%$ at the 
high $k$ (small scale) extreme and $\sim 85\%$ at the low $k$ (large scale) one. 
This indicates a clear advantage of this algorithm for applications in multi-dimensional data sets, due to its hyper-plane segregation and more particularly due to the non-linear kernel (radial basis function).

The degradation of results with the increase of dimension of the emulation problem is a consequence of the so-called curse of dimensionality \citep{bellman1957dynamic}, affecting most optimization, and thus also supervised learning, methods. This problem translates the need of increasing the size of the training set exponentially when adding dimensions in the feature space so that similar performances may be maintained as compared to the lower-dimensional case.
In this work, our choice of computer platform limits the augmentation of the  training set for emulations with two parameters to a factor of only four times the size of the training set of our one free parameter case. To maintain the high degree of accuracy of the single parameter case, we should further increase the size of the training dataset.

\begin{figure}[h!]
	\centering
    \includegraphics[width=\columnwidth]{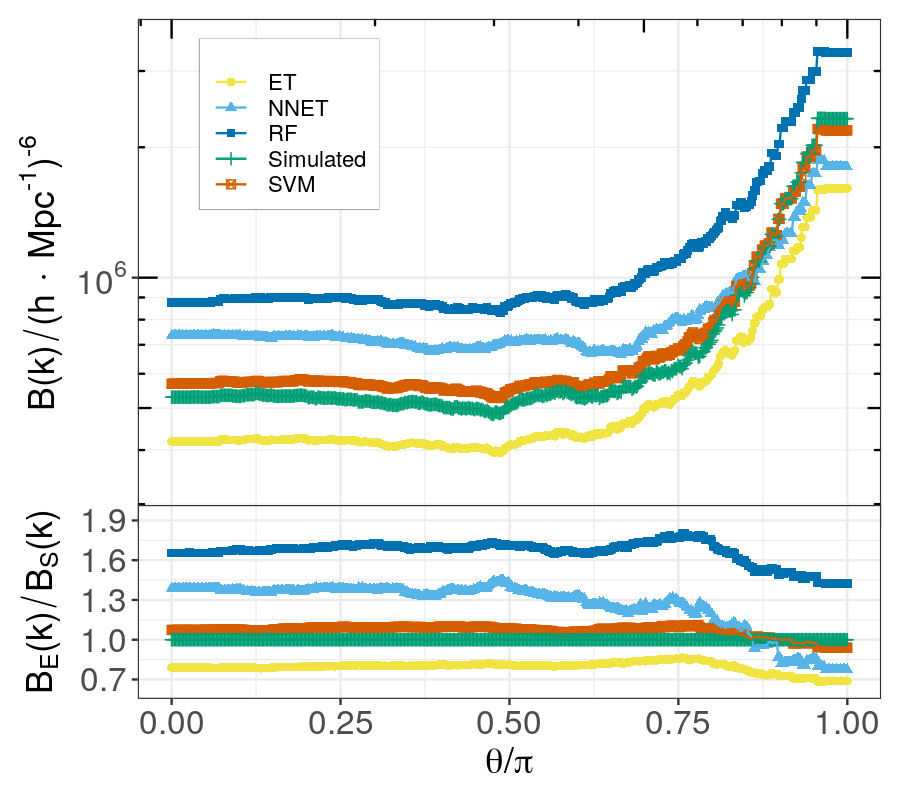}
	\caption[Bispectra of the Estimated Density Fields (2D case)]{The Bispectra of the estimated Density Field using the four algorithms on the two free-parameter ($\Omega_m$,z) at $\Omega_m=0.31$, and $z=0.5$. {\bf Top Panel:} Bispectra Curves. {\bf Bottom Panel}: Bispectra Ratio ($P_E(k)=$ Estimated;$P_S(k)=$ Simulated).}
	\label{fig:BiPowerAll2p}
	\centering
\end{figure}

Our Bispectrum results are shown in Fig.~\ref{fig:BiPowerAll2p}. 
A striking feature to notice is that all algorithms except ET overestimate the bispectrum, something that does not happen in the one-parameter emulation scenario. 
Since ET is the only algorithm which only considers one of the independent variables, redshift or $\Omega_m$, the other algorithms may be overestimating the bispectrum due to a degeneracy between both parameters. Indeed, the self-similarity of cosmological structure induced by gravity only simulations leads to some degree of degeneracy between the redshift epoch and $\Omega_m$ for a given scale of observation. A snapshot at high redshift with a given $\Omega_m$ displays a similar distribution of densities as a lower redshift snapshot with a lower $\Omega_m$. This is an approximate argument that does not take into consideration the impact of $\Omega_m$ on the epoch of transition from matter to cosmological constant domination (which happens at different $z$ for different $\Omega_m$), but generally applies to simulation snapshots away from these transitions redshifts.   

These results further indicate that SVMs are the best method among those studied in this work, since as we can see, even looking at higher order correlations it still out-performs the other algorithms. It is the only algorithm able to maintain an accuracy below $15\%$ for the entire $\theta$ domain. 

\section{Discussion}
\label{sec:disc}
In this section, we discuss some of our key findings. We first discuss the computational efficiency of our methods, as measured by CPU run times. Subsequently, we analyze the algorithms employed in our investigation, providing a detailed assessment of their performance and potential for application in related research. Finally, we conclude by outlining possible avenues for improvements in our approach.

\subsection{Efficiency}
In Table \ref{tab:DMemulTimes} we present the CPU run times for the best-performing algorithms in both emulation cases considered in this work. These cases correspond to: (1) considering $\Omega_m$ as a single free parameter in the first row of the table, and (2) using both $\Omega_m$ and $z$ as free parameters, in the second row. For the single free parameter case, we present the results trained with the dataset of 23 simulation snapshots at z=0. The run times shown in the Table correspond to both the PC estimation and to the de-projection into the final emulated density field, and we also present the sum of the two as the "Total emulation time" column.

\begin{table}[h!]
\caption[N-body Density Field Estimation CPU Running Times]{CPU run times for the emulation step our proposed N-body Density Field emulation pipeline.}
\resizebox{\columnwidth}{!}{  
\begin{tabular}{|c|c|c|c|}
\hline
\textit{\textbf{Emulation}} &
  \textbf{PC Estimation} &
  \textbf{De-projection} &
  {\textbf{Total emulation time}} \\ \hline
\textbf{\begin{tabular}[c]{@{}c@{}}$\Omega_m$\\ (NNET)\end{tabular}} &
  0.077s &
  0.719s &
  {\textbf{0.796s}} \\ \hline
\textbf{\begin{tabular}[c]{@{}c@{}}($\Omega_m$,z)\\ (SVM)\end{tabular}} &
  0.082s &
  4.380s &
  {\textbf{ 4.462s}} \\ \hline
\end{tabular}
}
 \label{tab:DMemulTimes}
\end{table}

In terms of pure computational efficiency, the outcomes of our emulation pipeline seem highly satisfactory, particularly when considering that an N-body simulation of a similar scale requires 84 minutes to complete using the same system used for our emulation experiments. As exemplified in Table \ref{tab:DMemulTimes}, the total emulation times for the single and two free-parameter scenarios were in the order of 0.8 and 4.5 seconds, respectively. This represents a substantial improvement in efficiency, as our proposed approach is three orders of magnitude faster than the corresponding N-body simulation run time.

\subsection{The Algorithms}
The supervised learning algorithms used in this study warrant some discussion, as some noteworthy observations can be made. 

First, our results indicate that the Random Forest algorithm was successful at reproducing the statistical properties of the density fields, particularly the non-Gaussian features, as evidenced by the Bispectrum results in Fig.~\ref{fig:Bispecallz1p} for the single free-parameter case. However, the opposite behavior is observed in the two free-parameters scenario, where Random Forests significantly underperforms in comparison to the other methods. In contrast, the Extremely Randomized Trees algorithm yields consistently disappointing results across both parametric contexts, emerging as one of the poorest performers, which we find surprising given its conceptual similarity to Random Forests.

Second, the NNET and SVM algorithms stand out, as both demonstrate superior performance in their respective contexts. NNET yields the best results for the one-dimensional N-body regression case, as depicted in Figs.~\ref{fig:PowerAll1p} and \ref{fig:Bispecallz1p}. Meanwhile, SVM outperforms all other methods in the multi-dimensional regression context. Considering that our current cosmological models incorporate multiple parameters beyond those examined in this first study (e.g., Hubble constant, Baryonic Matter Density, power spectrum normalization, spectral index), SVMs seems to be a promising method in the more realistic high-dimensional context.

However, these results do not yet allow for definitive conclusions regarding the potential of these supervised algorithms. It is important to keep in mind that our hyper-parameter optimization process may not have achieved a global minimum in the regression error metrics and the algorithm cost functions. As our simple search was restricted to a constrained and finite range of hyper-parameter values, there is a possibility that the algorithms were confined to a local minimum, with further potential for improvement through a better and larger exploration of the hyper-parameter space. Nonetheless, our findings offer some indications of the potential of the analyzed algorithms.

\subsection{Future Prospects}
There are several potential avenues to improve and/or extend the methods used in this study. Some of these improvements could be implemented without significant modifications to the existing pipeline. For example, one option would be to expand the parameter space to include additional cosmological parameters such as the Hubble constant or spectral index. Naturally, this will also require an increase in the training set to sample the parameter space. Alternatively, the dark matter density fields could be emulated with the inclusion of hydrodynamics, which would similarly require a substantial increase in training data. While hydrodynamics are known to introduce highly non-linear effects that may be difficult to reproduce, the overall structure of the pipeline should remain very similar, perhaps only requiring the adoption of more advanced dimensionality reduction methods.

On the other hand, methods similar to those proposed here could also be coupled directly within N-body codes, perhaps to accelerate certain aspects of the simulation, such as certain scales of the simulation. This would naturally require more extensive modification to the pipeline and of the N-body codes.

Finally, an interesting application of our proposed methods that we are currently investigating involves attempting to emulate the hydrodynamics directly from dark matter-only simulations. This requires some fundamental changes on the proposed pipeline, but this exciting prospect would enable the production of emulated hydrodynamical N-body density fields with computational overheads constrained purely by the computational requirements of gravity simulations, which are naturally significantly smaller than those also considering hydrodynamics.

\section{Conclusions}
\label{sec:conc}
In this work, we presented a methodology for fast and accurate emulations of full N-body dark matter density fields based on dimensionality reduction and supervised learning. In this first study, we applied our methodology in $128^3$ density cells in a box of 100 $h^{-1}$Mpc, using cosmological parameters as free parameters. We adopted four supervised machine learning methods, namely Random Forest, Extremely Randomized Trees, Support Vector Machine and Neural Networks, in two different emulation scenarios using dark matter density and redshifts as free parameters.

In the first scenario, we performed emulations solely based on $\Omega_m$. The proposed emulation pipeline was able to accurately reproduce the Power Spectrum of the corresponding simulated field with less than a $1\%$ difference for $k>0.3h \, \textrm{Mpc}^{-1}$ at all redshifts. Furthermore, we achieved similar results when reproducing the Bispectrum for all $\theta$ values, with less than a $3\%$ difference for $z=0$ and less than a $2\%$ difference for the remaining redshifts. Our method was trained using 23 N-body simulations within the $\Omega_m$ range of $[0.05,0.6]$ with a step size of $\Delta \Omega_m=0.025$, and we evaluated the emulations using $\Omega_m=0.31$ in this scenario.

In the second scenario, the emulator was extended to include the redshift as a free parameter, thus slightly increasing the parameter space dimensionality. The emulator was able to achieve good results using a Support Vector Machine, with the Power Spectrum being reproduced to less than a $5\%$ difference for $0.3h\,\textrm{Mpc}^{-1}\leq k \leq 2h\,\textrm{Mpc}^{-1}$ and the Bispectrum to less than a $15\%$ difference for all $\theta$. For this scenario, our emulator was trained on simulations with the same $\Omega_m$ range as the previous scenario, but at four different redshift snapshots ($z=0$, $z=0.25$, $z=1$, $z=10$), resulting in a four-fold increase in the size of the training dataset. We evaluated the emulations against a simulation with $\Omega_m=0.31$ at redshift $z=0.5$.

In both scenarios, our dark matter density field emulation strategy presented percent-level accuracy compared to the original N-body simulation while requiring a significantly smaller amount of CPU resources: the emulations of the full density field can be performed in about $\sim$0.8 seconds when using $\Omega_m$ as a free-parameter, and $\sim$4.5 seconds when using $\Omega_m$ and the redshift as free-parameters, corresponding to three orders of magnitude improvement compared to the time required to perform a full N-body simulation at the same scale. This scale of computational efficiency with the preservation of accuracy in the resulting density field can open doors to a multitude of exciting studies that require a large exploration of the parameter space and that are hardly feasible with direct N-body simulations, including the possible adoption of  density fields directly within cosmological parameter inference pipelines and the exploration of alternative cosmological and dark matter models.

\section*{Acknowledgment}
This work was partially supported by the Portuguese Funda\c c\~ao para a Ci\^encia e a Tecnologia (FCT) through the Portuguese Strategic Programmes and the following Projects and Research Contracts: EXPL/FIS-AST/1368/2021, PTDC/FIS-AST/0054/2021, UIDB/FIS/00099/2020, UID/FIS/00099/2019, CERN/FIS-PAR/0037/2019, PTDC/FIS-OUT/29048/2017, PTDC/FIS-AST/31546/2017, IF/01135/2015, and SFRH/BPD/74697/2010. 
AKM additionally acknowledges the support of the Caltech Division of Physics, Mathematics and Astronomy for hosting research leaves during 2017-2018 and 2019, when some of the ideas and codes underlying this work were initially developed. Finally, the authors thank Arlindo Trindade for useful discussions and Rafael S. de Souza for the useful comments during the defense of the Master thesis of MC and which also helped to improve this work.

%
%
\bibliography{bibliography}

%

%
%
%
%

\end{document}